\documentclass{article}

\PassOptionsToPackage{sort,square,comma,numbers,compress}{natbib}

\usepackage[preprint]{neurips_2021}

\usepackage[utf8]{inputenc} 
\usepackage[T1]{fontenc}    
\usepackage{hyperref}       
\usepackage{url}            
\usepackage{booktabs}       
\usepackage{amsfonts}       
\usepackage{nicefrac}       
\usepackage{microtype}      
\usepackage{xcolor}         
\usepackage{graphicx,wrapfig}

\usepackage[linewidth=1pt]{mdframed}
\usepackage{tikz-dependency}

\usepackage{graphicx}
\usepackage{color}
\usepackage{wrapfig}
\usepackage{listings}
\usepackage{graphbox}
\usepackage{caption}
\title{Social Construction of XAI:\\ Do We Need One Definition to Rule Them All?}

\author{Upol Ehsan \& Mark O. Riedl\\
Georgia Institute of Technology, Atlanta GA, USA\\
\texttt{\{ehsanu,riedl\}@gatech.edu
\vspace{-10pt}}
}

\begin{document}

\maketitle

\begin{abstract}
There is a growing frustration amongst researchers and developers in Explainable AI (XAI) around the lack of consensus around what is meant by `explainability'. 
Do we need one definition of explainability to rule them all?
In this paper, we argue why a singular definition of XAI is neither feasible nor desirable at this stage of XAI's development. 
We view XAI through the lenses of Social Construction of Technology (SCOT) to explicate how diverse stakeholders \textit{(relevant social groups)} have different interpretations \textit{(interpretative flexibility)} that shape the meaning of XAI.
Forcing a standardization \textit{(closure)} on the pluralistic interpretations too early can stifle innovation and lead to premature conclusions. 
We share how we can leverage the pluralism to make progress in XAI without having to wait for a definitional consensus.

\end{abstract}

\section{Of Bicycles \& Explainable AI}
\vspace{-5pt}
A lack of consensus on the definition of explainability is often portrayed as a problematic aspect in Explainable AI (XAI). However, \textit{do we need one definition of XAI to rule them all?} Considering the current developmental stage of XAI as a field, we argue that \textit{it is neither feasible nor desirable to have a singular definition of XAI}. To understand why that is the case, we need to first talk about bicycles and analyze their evolution over the last 200 years through the lenses of Social Construction of Technology (SCOT), a theory in Science \& Technology Studies that explains how human action foundationally gives shape and function to technology (for a full treatise of SCOT, refer to~\cite{bijker1997bicycles})

The word `bicycle' may evoke a prototypical image in your head, one with two  wheels, pedals, a handlebar with brake levers, and a seat offset from the rear wheel. Despite being one of the most stable pieces of technology, did you know that bicycles evolved rather drastically? For instance, the first bicycles lacked brakes by design because young men of "means and nerves" wanted that thrill of controlling a dangerous object while rolling down the English countryside~\cite{bijker1997bicycles}. In the 1800s, the famous penny-farthing bicycle had a huge front wheel with a smaller back wheel. Many bicycles had fixed handlebars that prevented turning~\cite{hadland2014bicycle}. As we reflect on the evolution of the bicycle, why and how did things evolve the way they did?

We will address this question using three concepts from SCOT. First, we have \textit{relevant social groups}---stakeholders with skin in the game such as bikers, families of bikers, mechanics fixing bikes, etc. These are the ones who are involved in or affected by a technological development. Different relevant social groups have their own \textit{interpretive flexibility}--- interpretations of what it means to be a bicycle. Different interpretive flexibilities can give rise to different types of bicycles such as mountain bikes, electric bikes, BMX bikes, etc. Finally, we have the notion of \textit{closure}-- over time, some interpretations of the bicycle achieved stability while others withered out (e.g., equal sized wheels won out over differently-sized wheels). This is how social aspects can construct the technology (of bicycles).

How does SCOT help with XAI? Just like bicycles, XAI has its relevant social groups. Many communities of practice are involved in XAI. Let's consider two relevant social groups: the Natural Language Processing (NLP) and Computer Visions (CV) communities. Given each group has its own ways of knowing (epistemology), there is interpretive flexibility on how they operationalize the notion of explainability. 
For instance, in NLP question-answering, explanations are often of the form of additional text that justifies the ground truth answer~\cite{wiegreffe2021reframing}. 
In CV, object recognition can consider saliency maps that show how visual features correlate to a predicted label~\cite{selvaraju2017grad}.
Given different operationalizations, their evaluations can also be different. This is a major reason behind the current pluralistic notions of XAI. 
This is to be expected because, unlike bicycles, we don't have 200+ years of development to reach clusters of closures yet.

Even though we do not have full closures in our interpretations of explainability, as a community there is a general consensus that our shared goal is to make an AI system’s functioning or decisions easy to understand by people~\cite{liao2021human}. There is a growing communal awareness that explainability is more than algorithmic transparency. That is, solving XAI challenges may require more than just ``opening the black-box''~\cite{ehsan2021expanding}. For example, Human-centered XAI (HCXAI) advocates to tackle XAI problems through a sociotechnical view (vs. a purely technical one)~\cite{ehsan2020human}. We need to consider \textit{who} is opening the box just as much as the algorithmic mechanisms of opening it.  Different ``whos'' need different explanatory facets to understand the AI~\cite{ehsan2021explainable,dhanorkar2021needs}. Whereas a lot of initial focus was on developers and data scientists as end-users of XAI systems, there is a growing recognition that we need to accommodate a diverse set of end-users, especially non-AI experts~\cite{miller2019explanation,ehsan2019automated}. If we had forced closure on what we mean by explainability too early, a lot of the emerging work would not have the intellectual space to flourish. 

\section{Cycling Onward \& Making Progress in XAI}
\vspace{-5pt}
So what does all of this mean?

\begin{enumerate}
    \item XAI is pluralistic. Given the different epistemic cultures co-existing in the space, we cannot expect monolithic conformity at this stage. 
    \item Pluralism, however, does not mean that anything goes; in fact, it’s the opposite---we need to be precise in our articulation of what we mean by explainability when we communicate. Thus, instead of using the term at face value, whenever we write a paper, we should strive to justify how our conception of explainability satisfies some of the shared goals we have in the space. 
    \item We should use SCOT to understand the sociology of XAI---who is saying what, when, and why. To grasp the flavor of explainability in a given context, we need to pay attention to a relevant social group's interpretation of it and how that informs their operationalization.
    \item While the notion of XAI is in flux, we are fortunate to join the conversation at this stage. We have substantial agency in steering the discourse, a privilege we need to exercise responsibly.
\end{enumerate}

A SCOT-based view helps us understand the sociological dynamics behind the pluralism in XAI, empowering us to focus on what XAI can \textit{do} for us instead of obsessing over definitional supremacy. It affords progress while simultaneously refining our shared understanding of the concept. It does not mean that we are advocating for epistemological anarchy or definitional relativism. 

If AI research could come this far without a monolithic consensus on the meaning of `intelligence', XAI can make progress without being bogged down in a definitional tug of war.

\begin{ack}
With our deepest gratitude, we thank discussants at the Human-centered XAI workshop at CHI where the initial ideas were developed.
We are grateful to members of the Human-centered AI lab at Georgia Tech for their feedback on the ideas. 
We are indebted to Michael Muller, Vera Liao, Elizabeth Watkins, and Samir Passi for their feedback on the early discussions.
Special thanks to Rachel Urban for generously providing proofreading feedback. This project was partially supported by the National Science Foundation under Grant No. 1928586
\end{ack}

\bibliography{main}
\bibliographystyle{unsrt}

\end{document}